\documentclass[%
 aip4-2,
 amsmath,amssymb,
 reprint,%
 floatfix,
]{revtex4-2}

\usepackage{graphicx}% Include figure files
\usepackage{dcolumn}% Align table columns on decimal point
\usepackage{bm}% bold math
\usepackage[utf8]{inputenc}
\usepackage[T1]{fontenc}
\usepackage{mathptmx}
\usepackage{etoolbox}
\usepackage{gensymb}

\makeatletter
\def\@email#1#2{%
 \endgroup
 \patchcmd{\titleblock@produce}
  {\frontmatter@RRAPformat}
  {\frontmatter@RRAPformat{\produce@RRAP{*#1\href{mailto:#2}{#2}}}\frontmatter@RRAPformat}
  {}{}
}%
\makeatother
\begin{document}

\preprint{AIP/123-QED}

\title{Strain engineering of the electronic states of silicon-based quantum emitters}
% Force line breaks with \\
\author{A.~Ristori}
\affiliation{Department of Physics and Astronomy, University of Florence, Via G. Sansone 1, I-50019 Sesto Fiorentino (FI), Italy}%
\affiliation{European Laboratory for Non-Linear Spectroscopy (LENS), Via N. Carrara 1, I-50019 Sesto Fiorentino (FI), Italy}%Lines break automatically or can be forced with \\

\author{M.~Khoury}%
\affiliation{Aix Marseille Univ, CNRS, Universit\'e  de Toulon, IM2NP, UMR 7334, F-13397 Marseille, France}%

\author{M.~Salvalaglio}
\affiliation{Institute of Scientific Computing, TU Dresden, 01062 Dresden, Germany}%
\affiliation{Dresden Center for Intelligent Materials (DCIM), TU Dresden, 01062 Dresden, Germany}%

\author{A.~Filippatos}
\affiliation{Dresden Center for Intelligent Materials (DCIM), TU Dresden, 01062 Dresden, Germany}%

\author{M.~Amato}
\affiliation{Laboratoire de Physique des Solides, Université Paris-Saclay, CNRS, Orsay 91405, France}

\author{T.~Herzig}
\affiliation{Division of Applied Quantum Systems, Felix-Bloch Institute for Solid-State Physics, University Leipzig,
Linn\'estrasse 5, 04103 Leipzig, Germany}

\author{J.~Meijer}
\affiliation{Division of Applied Quantum Systems, Felix-Bloch Institute for Solid-State Physics, University Leipzig,
Linn\'estrasse 5, 04103 Leipzig, Germany}

\author{S.~Pezzagna}
\affiliation{Division of Applied Quantum Systems, Felix-Bloch Institute for Solid-State Physics, University Leipzig,
Linn\'estrasse 5, 04103 Leipzig, Germany}

\author{D.~Hannani}
\affiliation{Aix Marseille Univ, CNRS, Universit\'e  de Toulon, IM2NP, UMR 7334, F-13397 Marseille, France}

\author{M.~Bollani}
\affiliation{Istituto di Fotonica e Nanotecnologie-Consiglio Nazionale delle Ricerche,
Laboratory for Nanostructure Epitaxy and Spintronics on Silicon, Via Anzani 42, 22100 Como, Italy}

\author{C.~Barri}
\affiliation{L-NESS, Dipartimento di Fisica, Politecnico di Milano, 20133 Como, Italy}

\author{C.~M.~Ruiz}
\affiliation{Aix Marseille Univ, CNRS, Universit\'e  de Toulon, IM2NP, UMR 7334, F-13397 Marseille, France}

\author{N.~Granchi}
\affiliation{Department of Physics and Astronomy, University of Florence, Via G. Sansone 1, I-50019 Sesto Fiorentino (FI), Italy}
\affiliation{European Laboratory for Non-Linear Spectroscopy (LENS), Via N. Carrara 1, I-50019 Sesto Fiorentino (FI), Italy}

\author{F.~Intonti}
\affiliation{Department of Physics and Astronomy, University of Florence, Via G. Sansone 1, I-50019 Sesto Fiorentino (FI), Italy}
\affiliation{European Laboratory for Non-Linear Spectroscopy (LENS), Via N. Carrara 1, I-50019 Sesto Fiorentino (FI), Italy}

\author{M.~Abbarchi}
\affiliation{Aix Marseille Univ, CNRS, Universit\'e  de Toulon, IM2NP, UMR 7334, F-13397 Marseille, France}
\affiliation{Solnil, 95 Rue de la R\'epublique, 13002 Marseille, France}
\email{marco.abbarchi@im2np.fr}

\author{F.~Biccari}
\affiliation{European Laboratory for Non-Linear Spectroscopy (LENS), Via N. Carrara 1, I-50019 Sesto Fiorentino (FI), Italy}%Lines break automatically or can be forced with \\
\affiliation{Department of Physics and Astronomy, University of Florence, Via G. Sansone 1, I-50019 Sesto Fiorentino (FI), Italy}%

\date{\today}% It is always \today, today,

\begin{abstract}
Light-emitting complex defects in silicon have been considered a potential platform for quantum technologies based on spin and photon degrees of freedom working at telecom wavelengths. 
Their integration in complex devices is still in its infancy, and it was mostly focused on light extraction and guiding. 
Here we address the control of the electronic states of carbon-related impurities (G-centers) via strain engineering. By embedding them in patches of silicon on insulator and topping them with SiN, symmetry breaking along [001] and [110] directions is demonstrated, resulting in a controlled splitting of the zero phonon line (ZPL), as accounted for by the piezospectroscopic theoretical framework. 
The splitting can be as large as 18\,meV and it is finely tuned by selecting patch size or by moving in different positions on the patch. Some of the split, strained ZPLs are almost fully polarized and their overall intensity is enhanced up to 7 times with respect to the flat areas, whereas their recombination dynamics is slightly affected. 
Our technique can be extended to other impurities and Si-based devices such as suspended bridges, photonic crystal microcavities, Mie resonators, and integrated photonic circuits.   
%
% Please include a maximum of seven keywords
\end{abstract}
\maketitle

\section{Introduction} 
\label{sec:Introduction}

Light emitters in silicon (Si) based on complex impurities~\cite{davies2006,chartrand2018} are currently scrutinized for their applicability as photon and spin quantum-bits~\cite{khoury2022}.  
Their appeal is manyfold: 1) they are nominally identical and in ensemble emission they have a sharp zero phonon line (ZPL) featuring a broadening of about 10~$\mu$eV in conventional Si and less than 1~$\mu$eV (about 10$^{-3}$~nm) in spin-less, isotopically purified $^{28}$Si~\cite{chartrand2018}; 2) their ZPL lies below the Si band-gap, in the near-infrared range (1.15 to 1.45~$\mu$m), covering the telecommunication O, E, and S bands~\cite{lohrmann2017,baron2022}; 3) their recombination lifetime can be as short as a few ns~\cite{beaufils2018,durand2021b,Baron2022b}; 4) they are stable in temperature and time (no bleaching nor blinking)~\cite{redjem2020,durand2021b} and can be detected up 120\,K~\cite{khoury2022b}; 5) they have well-defined polarization axes~\cite{redjem2020,baron2022,durand2021b}. 

Beyond these intrinsic properties that are unmatched by their counterparts in other materials, the possibility to exploit them in quantum technologies is highly entrancing, provided the advantages that this solid-state platform offers. 
Silicon technology steps on all the nano-fabrication methods developed in the last 50 years for electronic devices and it is, by far, more advanced than the technology applied to any other material: fabrication of electronic components can be provided in an industrial production chain with high material purity (99.9999999$\%$, \textit{nine nines}), large wafers up to 17 inches, silicon on insulator (SOI) wafers up to 12 inches, \textit{p} and \textit{n} doping, top-down lithography with nanometric resolution (e.g. based on deep-UV and plasma etching), availability of isotopically purified $^{28}$Si wafers~\cite{lo2009,itoh2014,becker2006,maurand2016,MAZZOCCHI2019,sabbagh2019}. 

The renewed interest in this class of Si-based emitters in the context of quantum technologies resulted in several breakthroughs over the last few years with the demonstration of: 1) single-photon emission from a large zoology of well-known (e.g. G-, W-, T-centers) and unknown defects~\cite{redjem2020,hollenbach2020,baron2022,durand2021b,prabhu2022,Baron2022b}; 2) photon coalescence~\cite{komza2022}; 3) spin control~\cite{bergeron2020,bergeron2020b,higginbottom2022}; 4) integration in photonic devices, such as Mie resonators~\cite{higginbottom2022,khoury2022b,hollenbach2022b}, integrated photonic circuits~\cite{deabreu2022,prabhu2022,komza2022}, ring resonators~\cite{lefaucher2022}, and photonic crystals~\cite{errando2022,redjem2023} providing Purcell effect; 5) position control of the emitters with localized ion implant~\cite{hollenbach2022}; 6) coherent population trapping and Autler-Townes splitting~\cite{higginbottom2023}.  

Most of these works focused on defect creation, enhancement of light emission, extraction, and guiding, in order to better detect and manipulate these single photon sources. A further step to control their properties requires a precise tuning of the photoluminescence frequency and its polarization (e.g. for setting the coupling of the ZPL with a photonic resonance). 

In this paper, we show that a large and tunable splitting of the ZPL of G-centers can be simply obtained by etching square patches in a [001]-oriented SOI wafer followed by deposition of SiN via plasma-enhanced chemical vapor deposition (PE-CVD). Compressive and tensile strain is obtained by adjusting the plasma frequency during SiN deposition, as confirmed by Raman spectroscopy. 
Symmetry breaking of the silicon unit cell along the vertical [001] direction is obtained at the center of the square patches, whose size sets the magnitude of the strain and thus the corresponding ZPL splitting in two main lines. Moving from the center of a patch towards the suspended (leaning) part, the symmetry breaking occurs also along the $\langle 110 \rangle$ directions leading to a splitting of the ZPL in four lines. 
Finite element methods calculations of the strain coupled to the \textit{piezospectroscopic} theory~\cite{kaplyanskii1967} confirms the overall picture observed in photoluminescence measurements. Partial or total polarization of the split ZPLs accounts for the alignment of the emitting dipoles along specific crystallographic directions.
Embedding the emitters in these structures entails an intensity enhancement of their photoluminescence up to a factor of 7 with respect to the flat counterpart, which is ascribed to a larger extraction of light, as confirmed by time-resolved experiments and finite element simulations.

\section{Results}
\label{sec:Results}

\subsection{Sample fabrication}
\label{sec:Sample fabrication}

\begin{figure}[h!]
\centering
\includegraphics[width=\columnwidth]{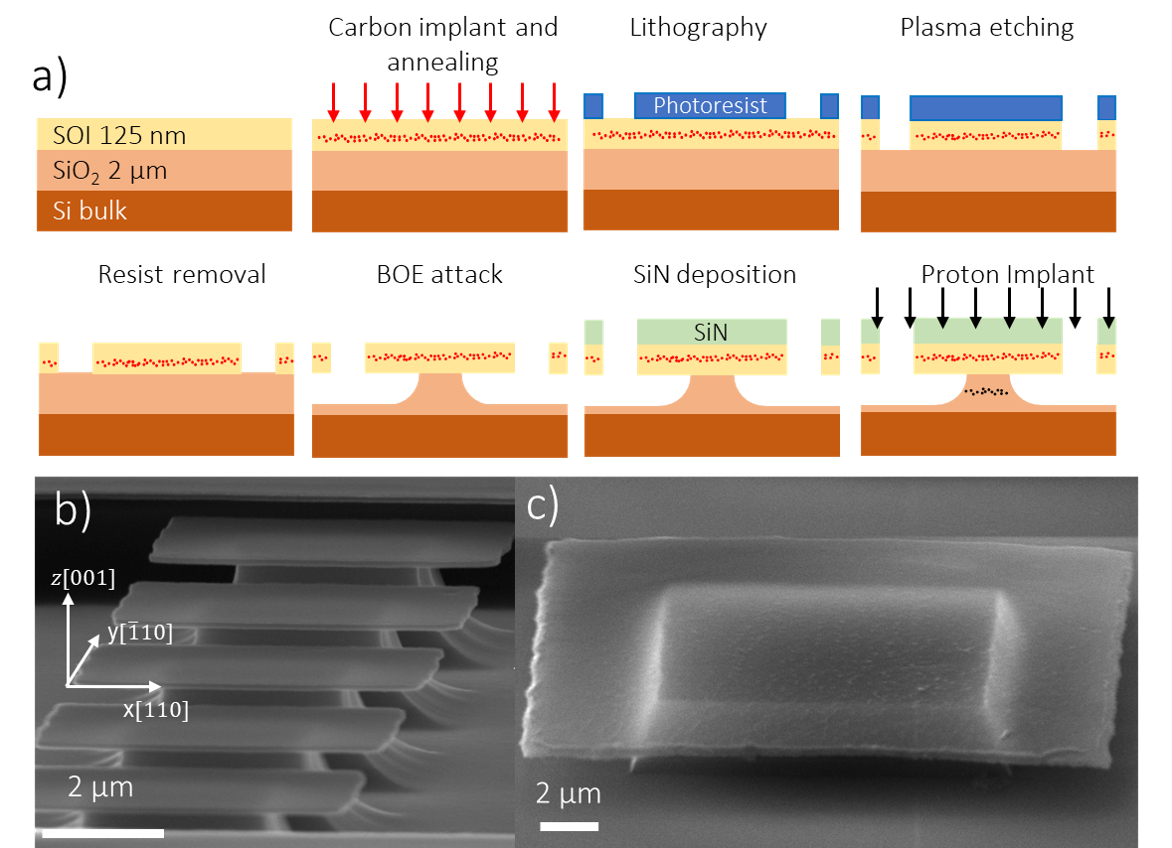}
\caption{ \textit{Sample fabrication}. a) A 125\,nm thick SOI sample is implanted with carbon ions and recrystallized by annealing at high temperature (1000~$^{\circ}$ C for 10 seconds in N$_{2}$ atmosphere). By optical lithography, square patches with variable side lengths are defined. The buried oxide underneath the SOI patches was partially removed via chemical etching using a buried-oxide etcher (BOE). SiN is deposited via plasma-enhanced chemical vapor deposition (PE-CVD). G-center activation is performed via proton implant. Details of the sample fabrication are provided in the dedicated section and in Table~\ref{tab:table1}. b) Scanning electron micrograph (SEM) of SOI patches before SiN deposition. c) SEM of a strained SOI patch with SiN atop.}
\label{fig:figure1}
\end{figure}

A detailed description of sample fabrication is provided in the dedicated section at the end of the paper (Section~\ref{sec:Experimental Section}). Here we refer to \textbf{Figure~\ref{fig:figure1}} where the main fabrication steps and some real examples are shown: a 125\,nm thick SOI is implanted with C ions and recrystallized by annealing. The underlying SiO$_{2}$ is partially removed with a diluted HF acid. The samples are then topped with SiN and etched  by optical lithography in square patches having different side length $L$.  Finally, proton implant is used to activate the emitting G-centers.  

In this work, we study three samples with etched patches. The details of their fabrication conditions are provided in Table~\ref{tab:table1}.

\begin{table*}[h!]
\caption{\label{tab:table1}%
List of samples and fabrication parameters: carbon implant energy and corresponding depth, and carbon dose; PE-CVD deposition conditions for SiN including high-frequency plasma time fraction (HF$\%$), high-frequency and low-frequency plasma power (HF, LF) corresponding type of strain (compressive or tensile, C or T), SiN thickness as measured by ellipsometry; proton implant energy and corresponding depth, and proton dose}
%\begin{ruledtabular}
\small
\addtolength{\tabcolsep}{0.38mm}
\begin{tabular*}{\textwidth}{c|cc|cccc|cc}
\hline
%\textbf{Sample} & \textbf{C ions implant} &  & \textbf{PE-CVD}  &  &  &  & \textbf{Proton implant} &  \\
\textbf{Sample} & \textbf{C ions implant} &  & \multicolumn{4}{l|}{\textbf{PE-CVD}} & \textbf{Proton implant} &  \\
\hline 
& Energy (Depth) & Dose & HF Plasma & Plasma Power & Strain & SiN & Energy (Depth) &  Dose \\
& keV (nm) & $10^{14}$\,cm$^{-2}$ & time \% & HF, LF (W) &
C, T & nm & keV (nm) & $10^{14}$\,cm$^{-2}$ \\
\hline 
%\colrule
A & 6 (25)   &  5.5  & 0   &   0, 100 & C & 108 & 90 (830) & 1\\ %1.3
B & 6 (25)   &  5.5  & 100 &  300, 0  & T & 102 & 90 (830) & 1\\ %1.4
C & 30 (100) &  6.5  &  70 &   20, 80 & T & 100 & 90 (830) & 1\\ %7.3 (7.2)
\hline 
\end{tabular*}
%\end{ruledtabular}
\end{table*}

\subsection{Finite element method simulations of strained patches}
\label{sec:FEMsim}

\begin{figure}[h!]
\centering
\includegraphics[width=\columnwidth]{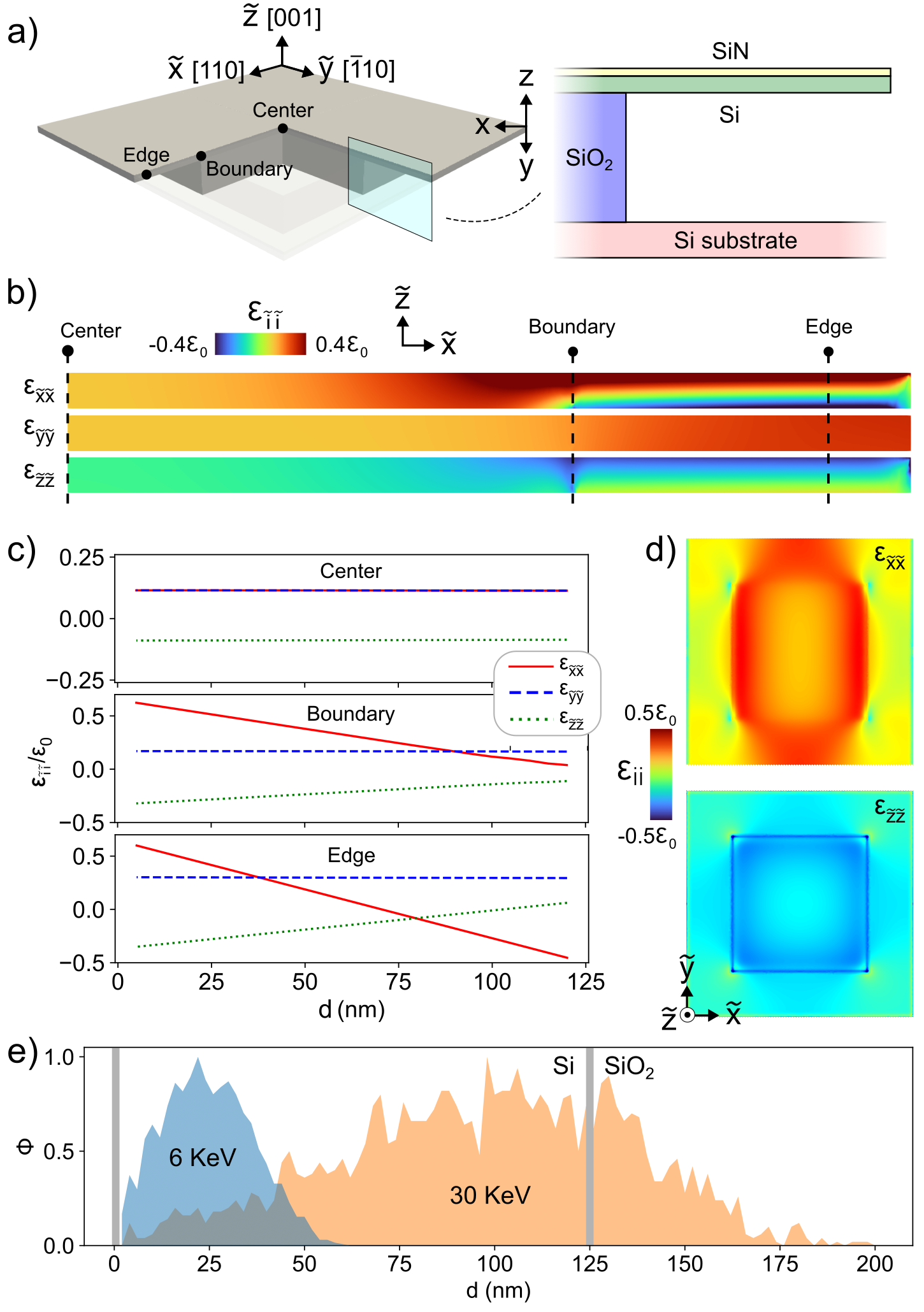}
\caption{\textit{Simulation of the elastic field}. a) Geometry in a perspective view (left) and 2D cross section highlighting the different materials (right). Three representative points are marked: ``\textit{center}", ``\textit{boundary}" between the leaning and suspended SiN/Si bilayer, and ``\textit{edge}" in the leaning part of the patch. b)
    Strain field in a ($\bar{1}10$) cross-section of the Si layer. Thickness is magnified ($\times 2$) for visualization purposes. c) Strain field components in the SOI layer as a function of the depth ($d$ corresponds to the $z$ direction [$00\bar{1}$]) from the SiN/SOI interface ($d$ = 0) at three $(x,y)$ points (center, boundary, and edge, as illustrated in panel a). d) Strain field component in the (001) plane at $d = 62.5$~nm.
    e) Normalized distributions of implanted carbon ions ($\Phi$) as a function of the depth in the SOI layer and in the SiO$_{2}$ BOX for two beam energies, namely 6~keV and 30~keV, obtained by SRIM software\cite{ziegler2010}.}
\label{fig:figure2}
\end{figure}

%Before addressing the effect of strain on the recombination dynamics of G-centers we provide a detailed description of the strain filed in the SOI patches topped with SiN. 
%A detailed description of the approach used is provided in Section~\ref{sec:Experimental Section}. We begin with a theoretical assessment of the deformation in the SOI patches topped with SiN

We provide a theoretical description of the elastic field in the strained patches via finite element method (FEM) calculations. Details of the method are provided in Section~\ref{sec:Experimental Section}. This analysis assesses the SOI deformation due to the action of the SiN stressor, thus allowing for the detailed explanation of the photoluminescence results, and independently complements the Raman spectroscopy results reported in the Supplemental Information provided at the end of this document. Without loss of generality, we focus first on a system where the length of the side of the membrane ($L$) is 10~$\mu$m (\textbf{Figure~\ref{fig:figure2}}). Moreover, we consider a SiN layer featuring a compressive in-plane deformation resulting in strain-field components $\varepsilon_{\tilde{x}\tilde{x}}$, $\varepsilon_{\tilde{y}\tilde{y}}$ and $\varepsilon_{\tilde{z}\tilde{z}}\equiv \varepsilon_{zz}$ with $\tilde{x}=[110]$, $\tilde{y}=[1\bar{1}0]$, and $z=\tilde{z}=[001]$ conveniently chosen to discuss effects on the Si crystal unit cell (while the canonical frame of reference is considered in the following when referring to $x$,$y$,$z$ directions).

For a flat film configuration, a tetragonal distortion of the stressor would occur with a linear out-of-plane displacement, no in-plane deformation, and no strain partitioning~\cite{ayers2016heteroepitaxy} (not shown), as actually measured by Raman spectroscopy for $L = 50~\mu$m (thus approaching the flat film case, see the Supplemental Information provided at the end of this document). 

For smaller membranes, additional free surfaces introduce extra elastic relaxation with strain partitioning among the stressor and the substrate, i.e. the Si layer. In particular, the SiN stressor induces an in-plane strain of the opposite sign in the Si layer and an out-of-plane strain with the same sign of the stressor in-plane strain, as observed in general for heterostructures exhibiting lateral free surfaces~\cite{Glas2006,Salvalaglio2014} (Figure~\ref{fig:figure2}b)). 

At the center of the structure ("\textit{center}" in Figure~\ref{fig:figure2}a),
%($\tilde{x} = 0;  \tilde{y} = 0$)
owing to the small relative thickness of the SiN/Si bilayer compared to their lateral size, a film-like configuration is obtained, and all the strain components vary negligibly along $z$, top panel) and $\varepsilon_{\tilde{x}\tilde{x}} = \varepsilon_{\tilde{y}\tilde{y}}$ (Figure~\ref{fig:figure2}c). As such, an isotropic in-plane deformation is achieved: in the plane of the patch the symmetry of the Si unit cell is preserved, whereas it is broken along $z$ under the action of $\varepsilon_{zz}$ that differs from the in-plane value. 

When moving towards the edge of the SiO$_2$ pedestal ("\textit{boundary}", Figure~\ref{fig:figure2}b)-d)), the crystal unit cells are deformed asymmetrically along $\tilde{x}$ and $\tilde{y}$, owing to different actions of free surfaces with normal along these directions (Figure~\ref{fig:figure2}b)). Similar asymmetric behavior is obtained in the free-standing part of the SiN/SOI bilayer. Here, significant variations of the strain values occur along the vertical direction (Figure~\ref{fig:figure2}c), central and bottom panels). 

The deformation close to the edges of the membrane can be interpreted in terms of the relaxation induced by a stressor on a suspended substrate with lateral free surfaces, known to lead to a bending of the bilayer ~\cite{Freund1999}. A bending downwards is expected for the relaxation of an initially flat bilayer with a compressive strain of the upper layer. 

Thus, both in-plane and out-of-plane symmetry-breaking is achieved when moving far from the center. This generally holds true over the whole square area of the membrane featuring additional relaxed region at the corners owing to the proximity of free surfaces with normal along both $\tilde{x}$ and $\tilde{y}$; see  $\varepsilon_{\tilde{x}\tilde{x}}$ and $\varepsilon_{\tilde{z}\tilde{z}}$ in the (001) plane at a depth of 62.5~nm in the Si layer in \textbf{Figure~\ref{fig:figure2}}d)  ($\varepsilon_{\tilde{y}\tilde{y}}$ would correspond to $\varepsilon_{\tilde{x}\tilde{x}}$ rotated by 90$^\circ$).
Therefore, we can conclude that properties depending on the symmetry breaking along the [001] crystallographic direction (i.e. the $z$ direction) are then expected to emerge everywhere. Instead, properties depending on the symmetry breaking along the $\langle 110 \rangle$ directions, are expected to emerge close to and at the suspended part of the patches. 

Note that the G-centers are placed at different depths $d$ along the vertical z-axis according to the implant energy, with a relatively broad distribution owing to the implant straggling (\textbf{Figure~\ref{fig:figure2}}e)). At the center of the patch, all the emitters experience the same strain field, irrespective of their implant depth, offering an ideal scenario for strain engineering of electronic states. Moreover, for the sizes considered in this investigation, the strain at the center of the membranes  is found to scale as $1/L^2$, so the strain can be well controlled with the lateral side. This information is also supported by finite element simulation of the strain field as a function of the patch side (not shown). However, emitters placed far from the center (close to the edge of the pedestal and in the suspended part) and at different depths would experience different strains. Proper weighting should then be considered to account for the effect on the electronic states. 
%Following theoretical analysis will focus on estimation for G-centers placed at the center of the membrane.

%The contribution of the deformation along the [001] direction, the symmetric contribution of $\varepsilon_{xx}$ and $\varepsilon_{yy}$ as well as the symmetry breaking when moving towards the edges, can be further appreciated from the plots in Figure~\ref{fig:figure2} d): it showcases the strain distribution along $z$ in the SOI at three representative points, namely at the center of the structure, at the edge of the SiO$_2$ pedestal, and close to the edge of thesuspended membrane. 

%As discussed above, the G-center implantation is not uniform along $z$ within the Si layer. The normalized distribution along z for both 6~keV (samples A and B) and 30~keV (sample C) as obtained by SRIM simulations are reported in Figure~\ref{fig:figure2} c).

The information gathered with these simulations is used as input for the \textit{piezospectroscopic} model used to evaluate the splitting of the ZPLs of the G-centers in membranes having different lateral sizes $L$. The details of the \textit{piezospectroscopic} model are provided in Section~\ref{sec:Experimental Section}, while the results are discussed in the following section.

\subsection{\label{sec:PL} Photoluminescence spectroscopy of strained G-centers}

We now address the effect of strain on the G-center ZPL by studying their emission from SOI patches having different $L$  and topped with SiN in different conditions. First, we study the emission from the center of patches (uniaxial symmetry breaking) and later the case of emitters far from the center (biaxial symmetry breaking). The simple case of flat SOI topped with SiN (not yet etched in small patches) is analyzed in detail in the Supplementary Information provided at the end of this document for samples A, B, and C, and compared to the unstrained case.

\begin{figure}[h!]
\centering
\includegraphics[width=\columnwidth]{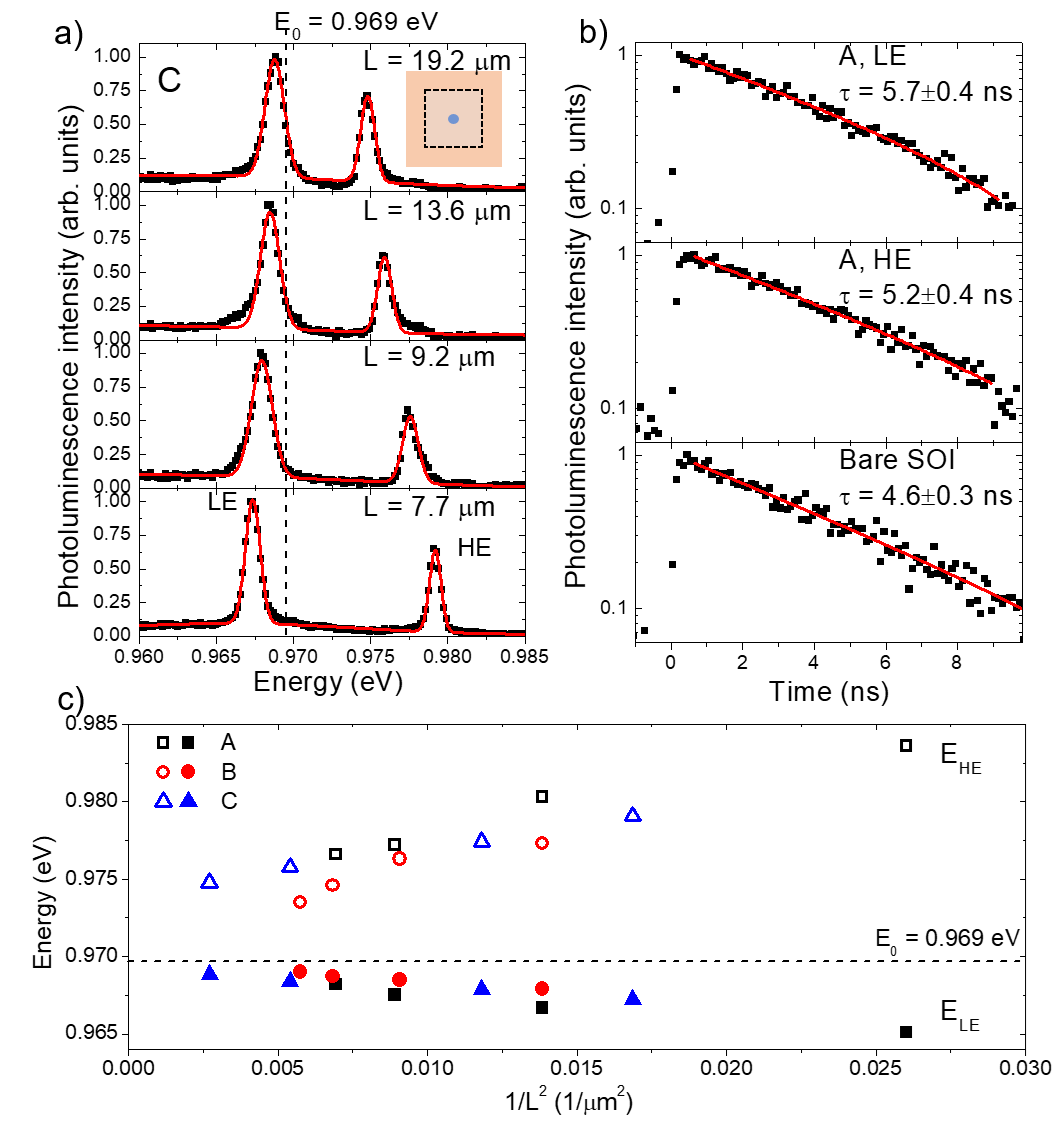}
\caption{\textit{Symmetry breaking along [001] direction}. a) Photoluminescence spectra collected at the center of the membrane for different square sides for sample C. The vertical dotted line highlights the energy of the ZPL emission in unstrained samples. The experimental data are represented as square dots whereas the red lines are Gaussian fits. The inset shows the scheme of the membrane highlighting the position of the excitation/collection spot. b) Time-resolved photoluminescence of the ZPL for unstrained SOI, HE and LE components from sample A. The experimental data are represented as square dots whereas the red line is a mono-exponential decay fit. c) Energy of HE and LE ZPL peaks for samples A, B, and C as a function of 1/L$^{2}$. The horizontal dotted line highlights the energy of the ZPL emission in unstrained samples.}
\label{fig:figure3}
\end{figure}

\begin{figure}[h!]
\centering
\includegraphics[width=\columnwidth]{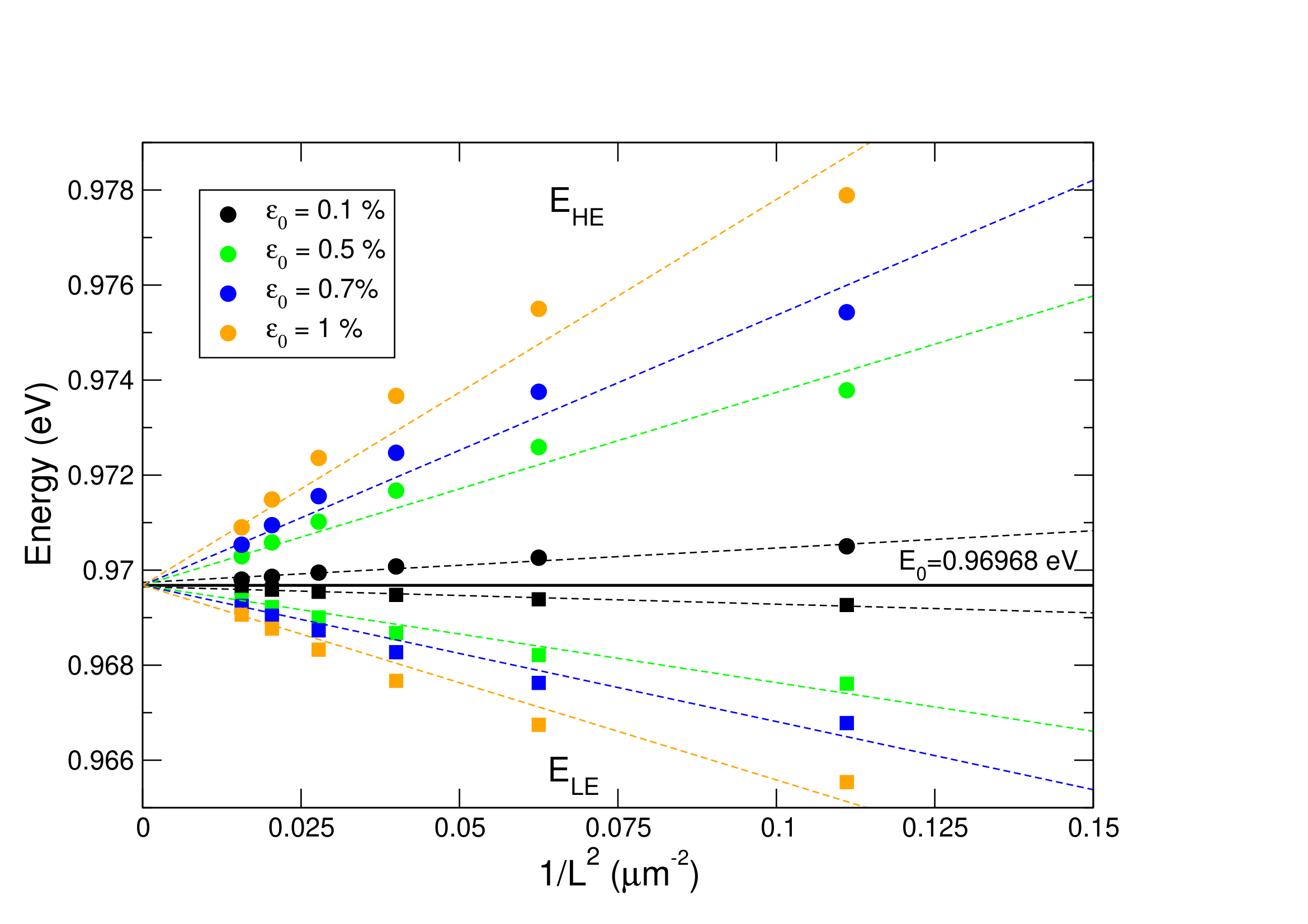}
\caption{ \textit{Theoretical energy shifts for [100] symmetry breaking}. Calculated $E_{\rm HE}$ and $E_{\rm LE}$ energy components at the central point (x = 0 and y = 0) for different values of the eigenstrain, $\varepsilon_{0}$ and at a depth of 25~nm, as a function of $1/L^2$, with $L$ the side length of the membranes.  The energy of the unstrained ZPL is highlighted. The data are derived by inserting in Eq.~\ref{energy_shift_100} the strain field simulated with FEM and the B$_{kl}$ coefficients from Ref.~\cite{tkachev1978}.  The dependence with $1/L^2$ was obtained by calculating $E_{\rm HE}$ and $E_{\rm LE}$ at the central point for $L = 6$ to 16~$\mu$m and then fitted for larger values. The trend at a depth of 100\,nm (not shown here) is similar. See more detail on the  \textit{piezospectroscopic} model that is described at the end of this paper in Section~\ref{sec:Experimental Section}.}
\label{fig:figure4}
\end{figure}

\textbf{Symmetry breaking along [001] direction}.  
Micro-photoluminescence experiments at the center of the patch showcase the typical behavior of G-centers under applied uniaxial stress~\cite{Foy1981} (\textbf{Figure~\ref{fig:figure3}a})). 
At this central point, as assessed by calculations illustrated in Section~\ref{sec:FEMsim}, %($\tilde{x} = 0$ and $\tilde{y} = 0$)  
the in-plane strain is isotropic (i.e. $\varepsilon _{xx} = \varepsilon _{yy}$ or, equivalently $\varepsilon _{\tilde x \tilde x} = \varepsilon _{\tilde y \tilde y}$), and the symmetry breaking of the Si cubic cell occurs only in the vertical direction [001] under the action of $\varepsilon _{zz}$ that differs from the in-plane strain. According to Ref.~\cite{kaplyanskii1967}, and Eq.~\ref{energy_shift_100}, a double peak appears in the spectrum with two lines having similar brightness and an energy separation $\Delta E_{\rm HE}+\Delta E_{\rm LE}$ that increases for smaller patch sizes.  The splittings in energy between the two new lines as a function of $1/L^2$ (with $L$ patch side) are approximately linear for all the samples (Figure~\ref{fig:figure3}c)) but with larger slopes for HE ZPLs and smaller slopes for LE ZPLs. Thus, the largest splitting is observed for the smallest patches and can be as large as 18~meV.

Time-resolved spectroscopy on both HE and LE ZPLs shows only small deviations of the recombination lifetime with respect to the unstrained case (Figure~\ref{fig:figure3}b)) that is in between 6 and 4~ns. These values are similar to those recently reported for individual G-centers in isotopically purified $^{28}$SOI~\cite{Baron2022b} and for ensemble in 220~nm thick SOI~\cite{beaufils2018}. Thus, the strain magnitude attained in these conditions does not modify the recombination dynamics, as also confirmed by the broadening of the HE and LE split lines (FWHM about 1.2~meV) which is not far from the unstrained case.

All this phenomenology is representative of all the investigated samples, irrespective of the nature and magnitude of the applied strain.

The energy shifts calculated at the patches center through the \textit{piezospectroscopic} theory (described at the end of this paper in Section~\ref{sec:Experimental Section}), using as inputs the strain field obtained by FEM calculations (see Section~\ref{sec:FEMsim}), confirm the almost linear dependence of $\Delta E$ with $1/L^2$ (\textbf{Figure~\ref{fig:figure4}}): HE and LE ZPLs shift with different slopes (larger for HE and smaller for LE) as a function of 1/$L^{2}$  as also measured on the patches and unlike previous reports on bulk samples~\cite{Foy1981,thonke1981}. 

From a quantitative point of view, the values of the strain field in the structure depend on the eigenstrain $\varepsilon_{0}$ (see also Section~\ref{sec:Experimental Section}), namely on the magnitude of the strain induced by the SiN layer. A reasonable agreement between theory and experiments can be found for $\varepsilon_{0}$ = 1$\%$ for the LE split ZPLs, which is consistent with the nominal strain expected by the fabrication procedure, whereas the measured HE slope is larger than the prediction. Several factors can lead to discrepancies between the \textit{piezospectroscopic} theory and the measured splittings: the model holds for bulk crystals and does not account for the complexity of the strain field in ultra-thin patches; measuring the photoluminescence at cryogenic temperature can further change the overall strain distribution with respect to the FDTD model reported here owing to the different thermal expansion coefficients of the SiN stressor, SOI patch, and underlying BOX pedestal. As such, a quantitative agreement between theory and experiments goes beyond the aim of this work.

%\subsubsection{\label{sec:PL [110]} Symmetry breaking along [001] and [110] direction}

\begin{figure}[h!]
\centering
\includegraphics[width=\columnwidth]{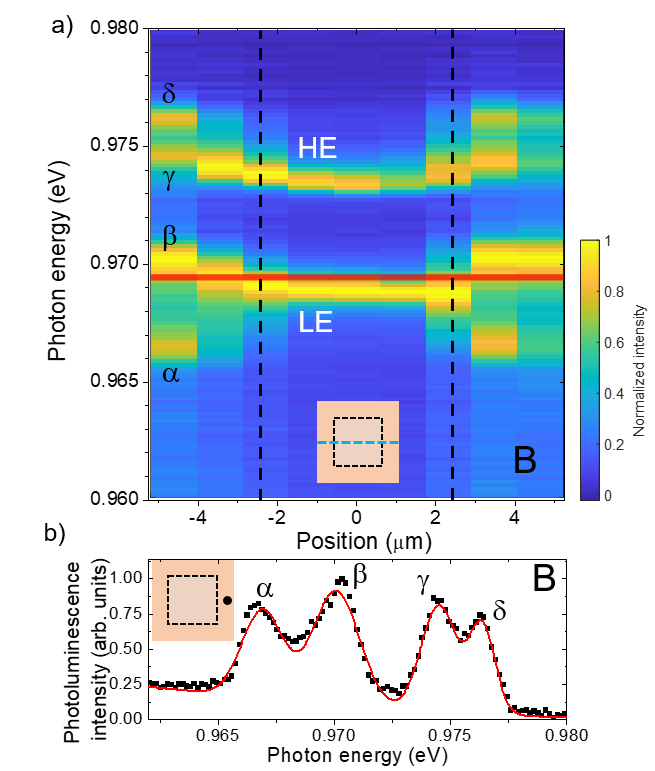}
\caption{ \textit{Symmetry breaking along [001] and [110] direction}.  a) Photoluminescence scan from patch side-to-side passing through the center (as highlighted in the inset). The horizontal, dashed lines highlight the edge of the SiO$_{2}$ pedestal. The vertical, red line highlights the ZPL energy of unstrained G-centers. b) Photoluminescence spectrum at the side of the patch. The experimental data are shown as squares, whereas the red lines are Gaussian fits. The insets show a scheme of the membrane and the position of the excitation/collection spot.}
\label{fig:figure5}
\end{figure}

\textbf{Symmetry breaking along [001] and [110] direction}. Moving from the center of the patch towards its side the Si unit cell is deformed differently along in-plane directions $\langle 110 \rangle$  (i.e. $\varepsilon _{\tilde x \tilde x} \neq \varepsilon _{\tilde y \tilde y}$) 
with still $\varepsilon _{\tilde z \tilde z} \neq 0$ (Figure~\ref{fig:figure2}). Moreover, depending on the position on the membrane and on the depth of the emitters (Table~\ref{tab:table1}),  the overall strain field applied to the G-centers at different depths can be very different (e.g. $\varepsilon _{\tilde x \tilde x}$ in the suspended part, at 25 and 100 nm, can change sign, \textbf{Figure~\ref{fig:figure2}} b)). 

This complex strain field across the patches is reflected in the micro-photoluminescence spectra, as shown in detail for sample B, with G-centers implanted at a depth of about 25 nm from the surface in \textbf{Figure~\ref{fig:figure5}}. In the inner part (not-leaning, between the vertical dashed lines in Figure~\ref{fig:figure5} a)), close to the center, the overall picture still resembles the properties of the central part (splitting in two lines of the ZPL). This points to a dominant role of symmetry breaking along the [001] direction with respect to other ones. When moving from the patch center, the two ZPLs undergo a spectral shift to higher energy under the action of the increasing strain. Moreover, owing to i) the onset of symmetry breaking in other directions, ii) the changes of the strain field across the SOI thickness, and iii) the straggling of the ion implant (\textbf{Figure~\ref{fig:figure2}}e)),  HE and LE ZPLs undergo a spectral broadening when far from the center (up to about 1.8~meV).  
%Notice that strain components vary differently along $z$ close of different free surfaces (see Figure~\ref{fig:figure2}).
%When moving from the center to the side, $\varepsilon _{yy}$ is rather constant along $z$ ($\varepsilon _{yy} / \varepsilon _{0} \sim + 0.2-0.3$) whereas the picture for the $x$ components at 25 ad 100 nm depth is very different with, respectively, $\varepsilon _{xx} / \varepsilon _{0}$ $\sim$+0.4 and $\varepsilon _{xx} / \varepsilon _{0}$ ranging in between $-0.1$ and $-0.3$. Also, the $z$ component changes a lot at 25 ad 100 nm depth, being respectively $\varepsilon _{zz} / \varepsilon _{0} \sim - 0.3$ and $\varepsilon _{zz} / \varepsilon _{0} \sim 0$. 
A clear splitting in four components, that is compatible with symmetry breaking along the [110] direction in addition to that along the [001], is obtained in the suspended part close to the edge of the membrane (see also Figure~\ref{fig:figure5}b)). 
A similar picture is found at the angle of the patches, in their suspended parts, where the ZPL is composite and splits into four main components (not shown).

\begin{figure}[h!]
\centering
\includegraphics[width=\columnwidth]{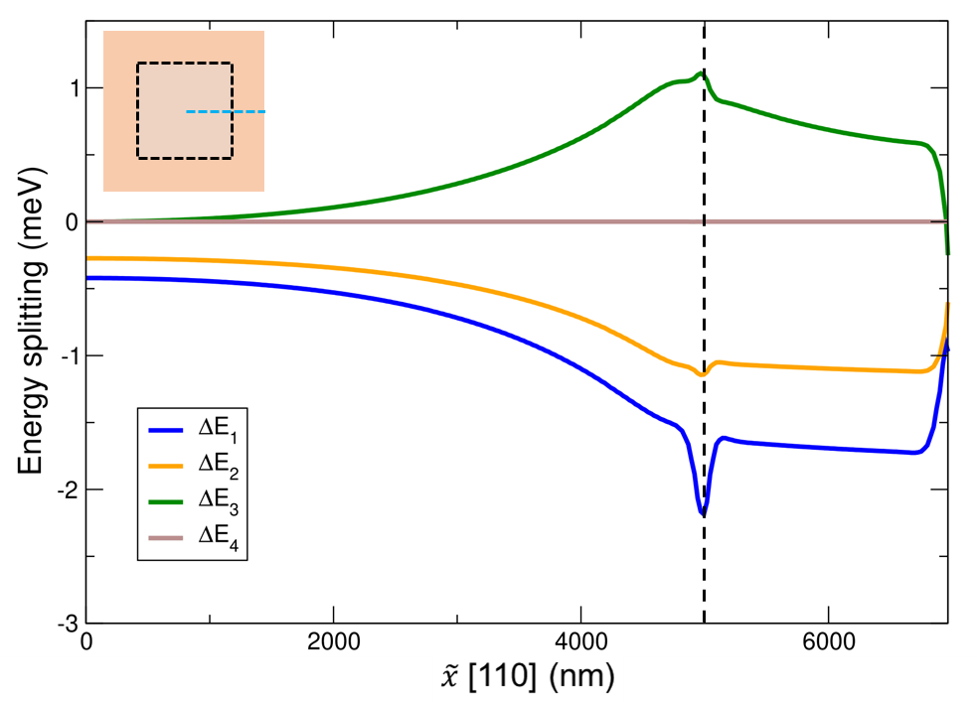}
\caption{ \textit{Theoretical energy shift for symmetry breaking along [001] and [110] directions}. The plot showcases the shift of the four, split ZPLs as a function of position from the patch center to the edge as obtained from the \textit{piezospectroscopic} model. We consider a patch having 14~$\mu$m side at a depth of 25\,nm with $\varepsilon_{0} = 0.3\%$. The vertical, black, dashed line highlights the end of the SiO$_{2}$ pedestal extending up to 5000~nm. The inset shows the scheme of the patch and the dashed, blue line highlights the position of the scan along the [110] direction considered in the model.}
\label{fig:figure6}
\end{figure}

As in the previous case, this symmetry-breaking picture is supported by the \textit{piezospectroscopic} model. Moving from the center towards the patch side, the strain field is not only composed of diagonal components, and the full \textit{piezospectroscopic} Eq.~\ref{energy_shift_complete} has to be considered. By inserting in this equation the simulated strain field and considering the four coefficients estimated in Ref.~\cite{tkachev1978}, one obtains different energy shifts (e.g. for a patch with $L = 14~\mu$m with C implant at a depth of 25\,nm, \textbf{Figure~\ref{fig:figure6}}): moving towards the side of the patch the number of components becomes four as observed in experiments. 

We observe that, using the full \textit{piezospectroscopic} model, at the patch center there are three split lines whereas in the experiments only two are visible. We interpret this as a combination of line broadening (much larger than the splitting), limited spectral resolution in experiments, and, potentially, different intensities associated with the split lines~\cite{Foy1981,thonke1981,kaplyanskii1967}.

\subsubsection{\label{sec:POLAR} Polarization degree of the zero phonon lines}

\begin{figure}[h!]
\centering
\includegraphics[width=\columnwidth]{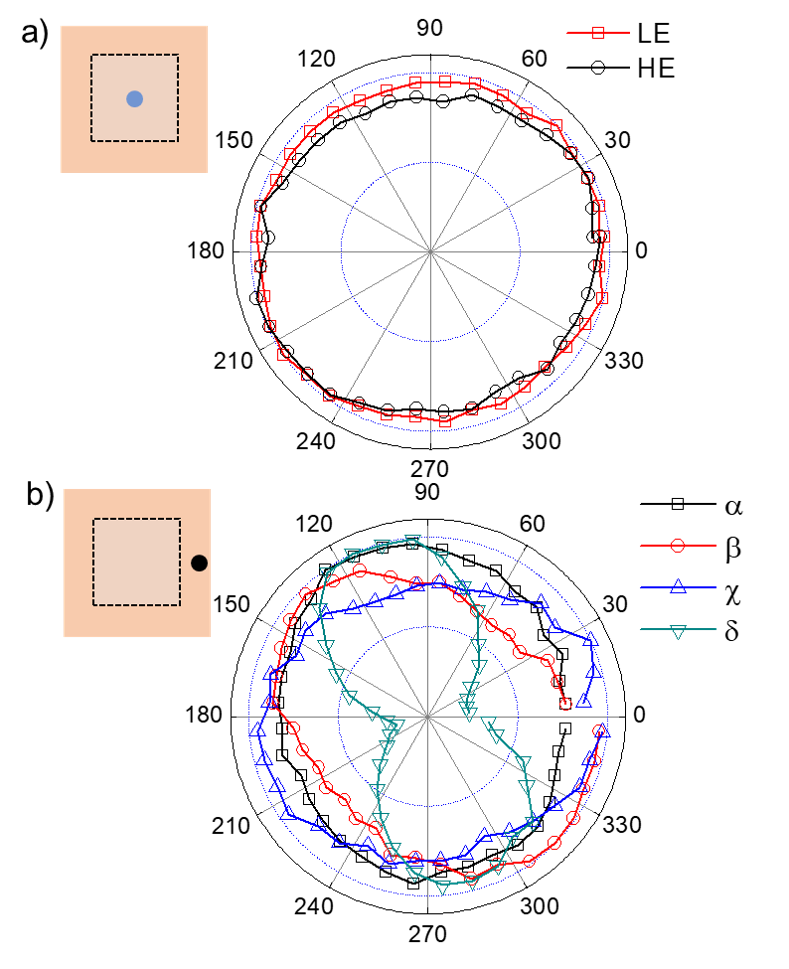}
\caption{\textit{Polarization-resolved photoluminescence}. a) Polar plot of the HE and LE ZPLs from the central point of a membrane. b) Polar plot of the $\alpha$, $\beta$, $\gamma$ and $\delta$ ZPLs from the side of a patch.  The insets show a scheme of the patch with the excitation/collection spot. All the data are relative to a patch with $L = 12.1~\mu$m on sample B.}
\label{fig:figure7}
\end{figure}

%\textbf{Polarization degree of the zero phonon lines}. 

Polarization-dependent micro-photoluminescence spectra of G-centers are registered for unstrained and strained samples (\textbf{Figure~\ref{fig:figure7}}). For unstrained G-centers (not covered by SiN), the ZPL is un-polarized, reflecting an even distribution of the emitting dipoles along the [110] and [1-10], in-plane directions (not shown) \cite{udvarhelyi2021}.  At the center of the membrane, the LE line shows a negligible degree of polarization, whereas, for the HE one, it  depends on the applied strain: in sample B, for $L = 13.2~\mu$m is about 3\%, for $L = 12.1~\mu$m is 8\% and for $L = 8.5~\mu$m is 50\% (here we only report the case of $L = 12.1~\mu$m). The polarization direction of the HE line is close to the [1-1-3] direction, that in our reference frame corresponds to +30 degrees (\textbf{Figure~\ref{fig:figure7}}a)). 

At the edge of the patch, in the leaning part, un-polarized to almost fully-polarized ZPLs (up to 70$\%$) are found, with the latter conditions typically observed for the high-energy component of the quadruplet $\delta$ that is oriented along the in-plane [12-1] direction (about 115$^\circ$ in our reference frame). Owing to a partial overlap of the split lines and their broadening, in these latter cases (e.g. as in Figure~\ref{fig:figure6}), it is not easy to extract the exact value of the polarization degree and our estimate is a lower bound. Similar results are observed at the corners of the membrane where the $\delta$  line has a polarization degree exceeding 50$\%$ (not shown).  These observations qualitatively agree with previous reports of polarized photoluminescence on strained G-centers ~\cite{Foy1981,thonke1981,kaplyanskii1967}.

\subsection{Photoluminescence enhancement}\label{sec:enhancement}

\begin{figure}[h!]
\centering
\includegraphics[width=\columnwidth]{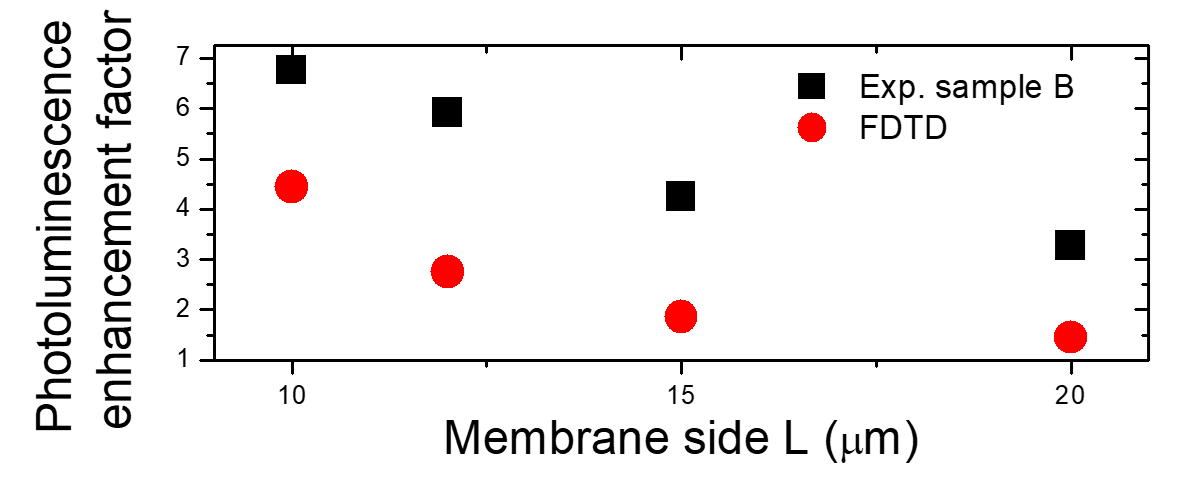}
\caption{\textit{Emission intensity.} Black squares: experimental emission enhancement from sample B. The data show the spectrally-integrated intensity of an individual spectrum collected at the center of the SiN/SOI membrane normalized by its counterpart from the flat areas (the collection spot is determined by the system lateral resolution and is about 1.1~$\mu$m). Red circles: FDTD simulations showing the spectrally integrated emission intensity normalized by its counterpart from the flat SOI.}
\label{fig:figure8}
\end{figure}

For sample B, we compare the emission intensity from the center of the patches with that from the flat SiN/SOI counterpart (black squares in \textbf{Figure~\ref{fig:figure8}}). We observe a clear trend showing an increased emission with a maximum enhancement of about a factor of 7 whereas, for larger values of $L$, the effect is reduced. 

Provided the negligible impact of strain on the recombination lifetime that cannot justify the observed enhancement (\textbf{Figure~\ref{fig:figure3}}b)) we look for a possible explanation in a photonic effect springing from the modified environment of the emitters in the membranes with respect to the flat SiN/SOI system. Through FDTD simulations (see a description of the simulation in Section~\ref{sec:Experimental Section}) we compare the emission of a dipole placed at the center of the patch with respect to the same dipole in an infinite SiN/SOI (red circles in \textbf{Figure~\ref{fig:figure7}}). The trend of the simulated data corresponds well to the one observed in the experiments, supporting the idea that the origin of the enhancement resides in an augmented light extraction springing from the finite size of the patches. 

Experiment and simulations display different enhancement factors (larger for experiments) that can be ascribed to the different conditions considered. Simulations consider only one dipole laying in the xy plane at a defined height whereas in rality we observe the emission from an ensemble of dipoles (excited by a laser spot and detected in a confocal configuration from about 1.1~$\mu$m), featuring several orientations and placed at slightly different depth in the SOI (\textbf{Figure~\ref{fig:figure2}}c)).

%\newpage 

\section{\label{sec:Discussion} Discussion}

The phenomenology presented here is similar to what was shown in early studies on G-centers ~\cite{tkachev1978,Foy1981,thonke1981,yukhnevich1965,yukhnevich1965a,yukhnevich1973} and, more generally, to other color centers that belong to the symmetry group of the first monoclinic type C$_{1h}$ in cubic crystals~\cite{baumann1967,kaplyanskii1967}. The photoluminescence splitting in a doublet that we find at the center of the patches mimics the effect of uniaxial strain along the [001] direction (no in-plane symmetry breaking). At the side of the patches, in the suspended part, where the Si cell symmetry is lifted also along the [110] direction, four split lines appear.  These experimental findings qualitatively agree with predictions obtained through the \textit{piezospectroscopic} model developed for strained cubic crystals~\cite{kaplyanskii1967}. 
%However, in the present case, strain engineering is obtained in an ultra-thin SOI, with a thin film coating while avoiding the use of bulk materials and stressors, as it was done so far~\cite{Foy1981,thonke1981}. This is an important difference that opens this method to realistic applications in quantum optics. %Moreover, the theoretical \textit{piezospectroscopic}  framework that we use here was developed for bulk materials and cannot quantitatively account for the observed splittings. In future studies, more refined models should be thus developed to have a better insight. 
We showed this control with a very simple micro-structure, avoiding the use of bulky samples and experimental setups making use of hydraulic pistons to apply stress to the samples~\cite{Foy1981,thonke1981}. It is also worth pointing out that, beyond the present proof of principle that was limited to relatively large patches ($L$ up to about 8~$\mu$m), the reduction of $L$ to about 1~$\mu$m m with conventional etching methods, or even below this value (e.g. exploiting solid state dewetting~\cite{khoury2022b,Toliopoulos_2020}), is possible. This could lead, potentially, to a stress that is more than one order of magnitude larger than what we show here, achieving strain regimes not attained so far. 

The importance of our results in the renewed context of quantum technologies with light-emitters in silicon is two-fold: 1) the possibility to control the splitting in a large range of energies (equivalent to a stress up to about 0.6~GPa~\cite{Foy1981,thonke1981}), allowing to tune the emission and, 2) the possibility to obtain, even from the ensemble, photoluminescence lines with a very large degree of linear polarization (as also reported in previous studies \cite{Foy1981,thonke1981,kaplyanskii1967}), demonstrating that an ensemble  of emitters shares the same energy and the same dipole orientation.
%(emission from ensemble of non-strained color centers showcases non-polarized light). Both observations are relevant for the exploitation of these emitters by integration in photonic devices and the production of quantum states of light.

The straggling of the ion implant might play a detrimental role for exploiting these emitters as quantum devices. On one hand, symmetry breaking along the [001] direction at the center of the membrane entails a uniform strain landscape in each of the three spatial directions. As such, whatever its position along z, an emitter will undergo the same splitting, shift, and polarization orientation as all the others. On the other end, for strain along the [110] direction (far from the center), the stress strongly depends on the position along the vertical z-axis: emitters created at different depth would undergo a different energy shift and splitting. Thus, in view of obtaining identical emitters (e.g. for quantum communication protocols), a reduction of struggling might be a hard requirement. This observation is also important for coupling the emitters to a localized photonic resonance~\cite{komza2022}. 
%
%For low-energy implant (e.g. sample A and B), the straggling is rather limited (all the emitters are within the first 50 nm from the surface, whereas for high energy (e.g. sample C), it is very relevant. 
%
Similar considerations hold for the high-temperature annealing step (required for re-crystallization after implant), which might spread the emitters over long distances along the vertical z-axis. However, in our case, where 1000 $^{\circ}$C annealing for only 20 seconds was used, we do not expect a relevant re-distribution of carbon. In fact, even for strong changes of the strain profile along the SOI thickness (e.g. at the patch sides), for low-energy implants at about 25~nm depth, where the straggling is rather limited and all the emitters are within the first 50 nm from the surface, the photoluminescence showcases rather sharp ZPLs: the measured line-widths in sample B at the edge of the patch is 1-2~meV, not far from 0.9~meV found in the flat, non -strained counterpart. These observations indirectly confirm that for low-energy implant annealing at 1000 $^{\circ}$C for 20 seconds does not affect too much the position of the implanted C ions in the SOI, unlike what was recently suggested based on theoretical simulations~\cite{komza2022}.

A potential element of concern for strained emitters could be the change in the underlying oscillator strength, impacting their brightness. Although a complete understanding of these features is beyond the aim of this work, we note that, from time-resolved photoluminescence experiments, the lifetime is weakly affected by the strain. Moreover, when monitoring the overall emission intensity, we observe larger intensity values from the patches with respect to the flat areas. For smaller patches, this enhancement is almost one order of magnitude and it might be even larger for smaller ones. 
%This is also in agreement with recent reports %of light extraction from sub-micrometric Si-%based antennas embedding T- and 
%G-centers~\cite{higginbottom2022,khoury2022b}.
This observation is important in the context of photoluminescence-intensity correlation with individual emitters (e.g. with Hanbury Brown and Twiss or Hong-Ou-Mandel interferometers), where light extraction is crucial for shortening the integration time of a measurement. In spite of its simplicity and lack of Purcell effect, our method which uses only low-resolution optical lithography, allows for a relevant increase in the collection efficiency.

The merit of our approach that relies on SOI lithography and deposition of a stressor film, is its compatibility with other devices, such as waveguides~\cite{prabhu2022,komza2022,deabreu2022,lefaucher2022} and Mie resonators~\cite{khoury2022b,higginbottom2022,hollenbach2022b}, that are relevant to all the other quantum emitters in Si~\cite{redjem2020,hollenbach2020,baron2022,durand2021b,prabhu2022,Baron2022b}. Our method can be generalized using suspended bridges providing uni-axial and bi-axial strain in arbitrary, in-plane directions~\cite{urena2013,suess2013,bollani2015,barget2016,gassenq2017}. Similar results can be obtained  by deposition of stressors~\cite{scopece2014} or by deposition of Si atop a strained substrate (e.g. made of SiGe)~\cite{david2020} opening the way to finely control the position of the ZPL of the emitters (e.g. with respect to a photonic resonance of a resonant cavity). For instance, this possibility is central for tuning several independent sources to the same energy in order to produce coalescent photons, as shown for III-V-based quantum-emitters~\cite{grim2019}.  In the same context, the evidence of an almost fully-polarized line measured on the ensemble is a promising signature for aligning all the emitting dipoles along the same axis. For unstrained G-centers in fact, the dipole orientation can be aligned along the crystallographic directions [110], [101], and [011], depending on the position of the self-interstitial Si atom~\cite{udvarhelyi2021} providing unpolarized light from ensemble emission. Our method suggests the possibility to engineer the polarization axes, providing at least one split ZPL with the same orientation for all the emitters.
%that is relevant to both coupling to a photonic mode in a resonator or in a waveguide and for the production of independent sources of coalescent photons.

%AVERAGING OF THE ANISOTROPY: Feofilov has termed this effect a ((latent optical anisotropy of noncubic centers [3]. FEOFILOV (P. P.), Physical Basic of Polarized Emission, Cons. Bureau, N. Y., 196

%"To reveal and to study the latent anisotropy
%of noncubic centers in cubic crystals, one should
%subject the crystal of interest to some anisotropic
%external action which would selectively affect indivi-
%dual groups of centers [l]. Since the physical proper-
%ties of centers are strongly anisotropic, the effect of a
%directional perturbation, and hence the response of
%centers to this perturbation will be different for
%centers of different orientation with respect to the
%direction of outer action."

Finally, the possibility to control the degeneracy of the electronic level in G-centers, and more generally in this class of impurities in silicon, might be relevant for the manipulation of their spin features. Symmetry breaking by application of strain could be the way to split the metastable triplet state of the G-center~\cite{O'donnell1983,ivanov2022} and thus encode spin-based quantum bits~\cite{arroyo2014,zu2014,foletti2009,wu2014}. The fast recombination time of G-centers with respect to similar emitters in Si (e.g. the T-center ~\cite{bergeron2020,bergeron2020b,higginbottom2022}) would provide a more efficient read-out of the spin degree of freedom providing a spin-photon interface.

\section{Conclusions}
\label{sec:Conclusion}

In conclusion, we showed that the splitting of the zero-phonon-line of the G-centers ensemble can be controlled producing doublets and quadruplets having large splittings, up to 18~meV, that is more than one order of magnitude larger than the broadening of the photoluminescence peaks. The splitting amplitude can be controlled by selecting a specific patch size. Symmetry breaking along [001] and [110] can be obtained respectively a the center of square patches and in their suspended parts. The merit of our approach is its simplicity, as it relies on conventional SiN deposition (chemical vapor deposition) and lithography (optical lithography and plasma etching). It can be further engineered by changing patch geometry and orientation with respect to the crystallographic axes, SiN thickness and, most importantly, it can be easily extended to most common photonic devices. It will be possible to apply the same principle to other color centers in Si and access the fine features of the underlying recombination dynamics allowing to exploit of the spin degree of freedom for quantum technologies.

%\begin{figure*}[bt]
%\centering
%\includegraphics[width=\linewidth]{FIGURES/FDTD Simulation.jpg}
%\caption{}%
%\label{fig: }
%\end{figure*}

%\begin{figure*}[h!]
%\centering
%%\includegraphics[width=\textwidth]{Figure Correction/Angle-resolved Spectroscopy.jpg}
%\caption{}
%\label{fig: }
%\end{figure*}

%\subsubsection{}

% \subsection{Discussion} \label{sec:Discussion}

%\subsection{Conclusion} \label{sec:Conclusion}

\subsection{Methods} \label{sec:Experimental Section}
%\textit{Measurement procedures}:
%\textit{Simulation} : The FDTD 
%\subsection{Second Level Heading}

\textbf{Fabrication of light-emitting, strained membranes}. The sample fabrication is described in \textbf{Figure~\ref{fig:figure1}} and the corresponding features are detailed in Table~\ref{tab:table1}.

The samples are obtained by dicing a 125~nm thick SOI laying atop a 2~$\mu$m thick buried oxide (BOX) on bulk Si (from Soitec) in 1~cm $\times$ 1~cm parts. They were first implanted with carbon ions at a depth of 25~nm (samples A and B) and 100~nm (sample C) following a well-established procedure~\cite{berhanuddin2012,beaufils2018}. After the implant, all the samples were flash-annealed in N$_2$ atmosphere for 20~s at 1000$^\circ$~C to cure the radiation damages. 

Each sample was patterned by optical lithography following this process: \\
- Chemical cleaning in a sonic bath, first with acetone and then with ethanol; \\
- Spin-coating of positive photo-resist (MICROPOSIT S1813) at 3000~rpm forming a layer of about 1.5~$\mu$m;\\ -Photo-lithography via UV laser system (Dilase 250 by Cloe). The pattern used for photoluminescence spectroscopy of strained G-centers consists of square patches whose size is tuned from about 15 to 5~$\mu$m; \\
- Resist development (using a solution from MICROPOSIT MF-319) to remove the parts that were not exposed to the UV laser; \\
- Pattern etching with a CF$_4$ plasma in a plasma-enhanced chemical vapor deposition chamber (PE-CVD, from Oxford); \\
- Sample cleaning with acetone; \\ 
- Partial etching of the BOX by immersion in an aqueous solution containing 10\% of Buffered Oxide Etch (BOE 10:1) to form partially-suspended membranes; \\
%
%An example of a patterned field containing the suspended membrane is shown in Figure~\ref{fig:figure1} GFGFG. 
%
- Deposition of a $\sim100~$nm thick SiN layer via PE-CVD.  Three depositions are performed using different parameters  changing the nature of the strain (tensile or compressive) applied to the SOI patches (Table~\ref{tab:table1}). This is done by tuning the time percentage of high- and low-frequency plasma (HF$\%$, which represents the fraction of time in which the high-frequency pulse is turned on with respect to the total pulse time expressed as a percentage). Changing HF$\%$ results in a compressive (${\rm HF}<60\%$), tensile ($\rm HF>65\%$) or almost unstrained ($\rm 60\%<HF<65\%$) SiN~\cite{Karouta_2012}. Ellipsometry (ESM-300 by Wollam) was systematically used to measure the thickness of the SiN layers deposited atop the SOI. The strained SiN affects the SOI patches resulting in a bent membrane, upwards for tensile strain, or downwards for compressive strain (\textbf{Figure~\ref{fig:figure1}}c)); 

- Implant with protons, to form the G-centers by inducing the interstitial Si~\cite{Berhanuddin_2012}. The beam energy was set to 90~keV in order to stop the protons within the BOX, avoiding the activation of unwanted emitters in the SOI or in the underlying bulk Si.

%\subsubsection{Raman spectroscopy setup}
\textbf{Raman spectroscopy setup}. Micro-Raman spectroscopy is performed at room temperature and is used to assess the stress in the flat samples and in the SOI patches. The setup consists of an Horiba-Jobin Yvon HR800-UV Raman spectrometer mounting a 1800~gpmm grating. It is characterized by a spectral resolution of about 0.5~cm$^{-1}$. The dispersed signal is then collected with a Jobin Yvon Synapse Si-based CCD camera. The samples are placed on a platform equipped with a $x-y$ translation stage to scan the sample surface and characterized by a spatial precision of about 250~nm. The samples are excited by a He:Ne laser ($\lambda=632.8$~nm) through a 50$\times$ objective lens (Olympus SLMPLN50X) with a numerical aperture NA~$=0.35$ mounted on a metallographic microscope (from Olympus). The Raman signal from the sample is collected via the same objective in a back-scattering configuration. 
%Thanks to a removable beam splitter it is possible to observe the sample surface with a CMOS camera. With the help of software it is possible to collect raster micro-Raman maps of the sample by synchronizing the movement of the sample with the spectra acquisition. To calibrate the system, before each measurement, a monocrystalline silicon sample is measured.

%\subsubsection{Ion implant of carbon and protons}

%TOBIAS, MARIO and ANDREA: refer to Figure 1 a)

%\subsubsection{Micro-photoluminescence spectroscopy setup}

\textbf{Micro-photoluminescence spectroscopy setup}. The micro-photoluminescence spectra  are detected with a custom-made confocal microscope setup in a backscattering configuration. The samples are placed in a low-vibration, He-flow cryostat (Janis ST-500), mounted on a  $x-y$ translation stage (Physik Instruments) to scan the sample surface with fine control on the position of about 250~nm. The samples' temperature is fixed at about $10~$K.

The excitation sources are a CW diode-pumped solid-state laser emitting at 532~nm (CNI MLL-III-532) and a mode-locked Ti:Sapphire tunable laser (Spectra Physics Tsunami, 700-900~nm spectral range, 200~fs pulse duration, 12.2~ns pulse period). 

The photoluminescence from the samples is collected by an infinity-corrected Mitutoyo 100$\times$ objective lens (NIR, $\mathrm{NA} = 0.7$), separated from the excitation by a dichroic mirror, spectrally dispersed by a spectrograph (Acton SP2300i) mounting a 600~gr/mm grating. The photoluminescence  is detected by an InGaAs array (Princeton Instruments OMA V-512) kept at $-100^\circ~$C by a cold finger immersed in liquid nitrogen. 

The spatial resolution of the micro-photoluminescence setup is about 1.1~$\mu$m, and the spectral resolution is about 350~$\mu$ eV.

Time-resolved PL (TR-PL) measurements were performed using the time-correlated single photon counting (TCSPC) technique using an InGaAs/InP APD (ID Quantique ID230) and a time correlator (ID Quantique ID900) interfaced with the PC. The time resolution of the system is about 200~ps.

Polarization measurements are performed by placing a rotatable half-wave plate and a fixed polarizer along the collection path.

%\subsubsection{Strain distribution by Finite element method simulation}
\textbf{Finite element method simulations of strain distribution}. The strain distribution in the system is computed by 3D Finite Element Method (FEM) calculations based on the linear elasticity theory ~\cite{bollani2015,landau1986theory} exploiting the commercial FEM package COMSOL Multiphysics. To model the experimental structure under investigation, we consider a square bilayer formed by a SiN layer of 100~nm atop a 125~nm thick silicon layer. This layer is supported at the center by a SiO$_{2}$ pillar with a square base. The whole structure is placed on a silicon substrate. The suspended part of the SiN/Si bilayer, from its edge to the edge of the SiO$_{2}$ pedestal, is 2~$\mu$m (independently of the lateral extension of the membrane). Simulations are performed for structures with a center-to-edge extension of the SiN on SOI ranging between 6 and 16~$\mu$m whereas the bottom SiO$_{2}$ (see \textbf{Figure~\ref{fig:figure2}}a), mimicking the geometry shown in \textbf{Figure~\ref{fig:figure1}}b) and c)). 

The elastic field is then computed by the FEM simulations to satisfy the mechanical equilibrium condition (without external forces), $\nabla \cdot \boldsymbol{\sigma} = 0$, the free surface boundary condition, $\boldsymbol{\sigma} \cdot \widehat{\mathbf{n}} = 0$ with $\widehat{\mathbf{n}}$ the normal to the free surface(s), and the Dirichlet boundary condition $\mathbf{u}=0$ with $\mathbf{u}$ the displacement field for the silicon substrate similarly to previous investigations of mechanical stress in heteroepitaxial systems~\cite{Salvalaglio2014,Bergamaschini2016}. 

The stress field is defined as $\boldsymbol{\sigma}=\mathbf{C}:(\boldsymbol{\varepsilon}-\varepsilon_0\mathbf{I})$ with $\boldsymbol{\varepsilon}=(1/2)(\nabla \mathbf{u}+(\nabla \mathbf{u})^{\rm T})$ the strain field and $\varepsilon_0$ the eigenstrain \cite{kinoshita1971elastic}, namely minus the mismatch of the SiN layer with respect to the Si layer induced by the growth process. $\varepsilon_0$ is found to vary according to the fabrication condition in the range $\pm 1\%$. The elastic constant tensor $\mathbf{C}$ is assumed to be isotropic for the purposes of this work. Expressed in terms of Young modulus (E) and Poisson ratio ($\nu$) they are: $E_{\rm Si}$=$130\,{\rm GPa}$, $\nu_{\rm Si}$=$0.27$, $E_{\rm SiN}$=$250\,{\rm GPa}$, $\nu_{\rm SiN}$=$0.23$, $E_{\rm SiO_2}$=$70\,{\rm GPa}$, $\nu_{\rm SiO_2}$=$0.17$.

We explicitly consider only the case of a SiN layer under compressive strain. However, the resulting strain distributions scale linearly with $\varepsilon_0$. Therefore, as we are interested in the ratios between different strain components, their distributions are provided in relative units w.r.t such an eigenstrain.

%It is worth stressing here that these simulations account for the presence of the top SiN and of the BOX pedestal. However, they do not consider the effect on temperature that might change the overall strain field owing to differences in the thermal expansion coefficients of SiN, Si, and SiO$_{2}$ (respectively in units of 10$^{-6} K^{-1}$, 1.4, 2.6, 0.5). However, at least at the center of the patch where the symmetry breaking is only along the vertical z-axis, the overall picture does not change, and the simulation can catch the main picture, as also indirectly confirmed by the agreement between simulations and experiments.      

\textbf{Theoretical \textit{piezospectroscopic} model}. The behavior of non-cubic color centers in cubic crystals can be described by adopting a \textit{piezospectroscopic} method~\cite{kaplyanskii1967,HughesPPS1966,thonke1981} coupling the observed energy splitting of the center spectral lines with the strain field. The transition energy shift in the emission of a center, $\Delta E$, can be written as: 
\begin{equation}
\Delta E =\sum_{k,l}B_{kl} \varepsilon_{kl}
\label{energy_shift}
\end{equation}
where $\varepsilon_{kl}$ are the strain field components (with $kl$ pointing at the canonical frame of reference), and $B_{kl}$ some coefficients which form a second rank symmetric tensor~\cite{kaplyanskii1967}. Both experimental~\cite{Foy1981,tkachev1978,ivanov2022} and first principles studies~\cite{ivanov2022,TimerkaevaJAP2018,udvarhelyi2021} have clearly demonstrated that G-centers, which consist of two substitutional carbons and one interstitial silicon atom, behave as first-type monoclinic centers (C$_{1h}$) with an axis along the [110] direction. In this case, the number of coefficients $B_{kl}$ is reduced to four, and Eq.~\ref{energy_shift} can be rewritten as: 
\begin{equation}
\Delta E = B_{1} \varepsilon_{zz} + B_{2} (\varepsilon_{xx} + \varepsilon_{yy})+2B_{3}\varepsilon_{xy}+2B_{4}(\varepsilon_{yz}-\varepsilon_{xz})
\label{energy_shift_complete}
\end{equation}
Depending on the particular strain field, Eq.~\ref{energy_shift_complete} can be further simplified. More specifically, in the case of a tensile or compressive deformation along the [100] direction, the number of split spectral lines is two, whereas it corresponds to four when the deformation is along the [110] direction. In the first case, Eq.~\ref{energy_shift_complete} becomes: 
\begin{equation}
\Delta E = B_{1} \varepsilon_{zz} + B_{2} (\varepsilon_{xx}+\varepsilon_{yy})
\label{energy_shift_100}
\end{equation}
Assuming that the energy separation $\Delta E$ is composed by an high energy shift, $\Delta E_{\rm HE}$, and a low energy shift, $\Delta E_{\rm LE}$ (corresponding to the two split spectral lines) Eq.~\ref{energy_shift_100} can be written as: 
\begin{equation}
\Delta E_{\rm HE} + \Delta E_{\rm LE}= B_{1} \varepsilon_{zz} + B_{2} (\varepsilon_{xx}+\varepsilon_{yy})
\label{energy_shift_100}
\end{equation}
if this equality holds, then:
\begin{equation}
\begin{split}
\Delta E_{\rm HE} &= E_{\rm HE}  - E_{0} = B_{1} \varepsilon_{zz}\\
\Delta E_{\rm LE}&=  E_{\rm LE}  - E_{0} = B_{2} (\varepsilon_{xx}+\varepsilon_{yy})\\
\end{split}
\label{energy_shift_100_EHE}
\end{equation}
where $E_{0}$ is the ZPL (0.969 eV). Eqs.~\ref{energy_shift_100_EHE} can be finally written as: 
\begin{equation}
\begin{split}
E_{\rm HE} &= B_{1} \varepsilon_{zz} + E_{0} \\
E_{\rm LE}&=  B_{2} (\varepsilon_{xx}+\varepsilon_{yy}) + E_{0} \\
\label{energy_shift_100_ELE}
\end{split}
\end{equation}
On the other hand, when the deformation is along the [110] direction, the non-diagonal components of the strain tensor cannot be neglected. In this case, the number of split lines is four, and Eq.~\ref{energy_shift_complete} has to be fully taken into account. 

We employed both versions of Eq.~\ref{energy_shift_complete} (full and simplified one) to theoretically evaluate the shift of the G-center ZPL moving from the center to the side of the patches. The strain field, $\varepsilon_{kl}$, is computed by 3D FEM calculations as described above while $B_{kl}$ coefficients are taken from previous works on [100] and [110] strained cubic Si~\cite{tkachev1978}. However, in our study, Eq.~\ref{energy_shift_complete} assumes a slightly more complex form. Indeed, as is shown in \textbf{Figure~\ref{fig:figure2}}, the strain field $\varepsilon_{kl}$ is a function of the distance from the center of the patch, $d$. As a consequence,  Eq.~\ref{energy_shift_complete} becomes:     
\begin{equation}
\begin{split}
\Delta E (d) = &B_{1} \varepsilon_{zz}(d) + B_{2} [\varepsilon_{xx} (d) + \varepsilon_{yy}(d)]~+\\
&+~2B_{3}\varepsilon_{xy}(d)+2B_{4}[\varepsilon_{yz}(d)-\varepsilon_{xz}(d)]
\label{energy_shift_complete_d}
\end{split}
\end{equation}
and, in the same way, Eqs.~\ref{energy_shift_100_ELE} can be written as: 
\begin{equation}
\begin{split}
E_{\rm HE} (d) &=  B_{1} \varepsilon_{zz} (d) + E_{0} \\
E_{\rm LE} (d) &=  B_{2} [\varepsilon_{xx} (d)+\varepsilon_{yy} (d)] + E_{0} 
\label{energy_shift_100_ELE_d}
\end{split}
\end{equation}
In this way, we build a predictive model to prove that the measured ZPL energy shifts are due to the difference in the strain field between the center and the side of the patches. Moreover, such a model can also demonstrate that the dependence of $\Delta E$ on $1/L^2$ (where $L$ is the patch side) is linear.        

\textbf{Finite Difference Time Domain simulations of light emission}. Calculations of light emission from the etched patches were performed by Finite Difference Time Domain (FDTD) method employing the commercial software Lumerical. %[Lumerical 3D Electromagnetic Simulator, https://www.lumerical.com/products/fdtd/]. 
To simulate the emission from a single G-center, an x-polarized, spectrally narrow, light-emitting electric dipole is used, centered around 0.970~eV with a pulse length of 3000 fs. We simulate two kinds of systems: 1) an infinite multilayer structure composed by (from top to bottom) a 100~nm thick SiN  layer, a 125~nm thick SOI layer, and a 2~$\mu$m thick SiO$_{2}$ layer atop a Si bulk. 2) Square patches of different sides are modeled analogously to the one used in FEM simulations described before, except made for the strain effect which is not included in these FDTD simulation domain. In both cases (infinite multilayer and finite size membranes) the dipole is positioned at 25~nm from the upper surface of the SOI layer (at the center of the membrane for the finite patch cases). A 2D monitor geometrically configured to reproduce an acquisition of NA = 0.7, is positioned atop of the multilayered structures and collects the signal coming from the system. The intensity is then calculated by integrating the emission over the monitor area (12~$\mu$m $\times$ 12~$\mu$m). For the sake of thoroughness, we also studied different configurations by changing the monitor distance from the multilayer and adjusting the monitor size in order to reproduce the experimental NA of 0.7. No significant differences were highlighted, and this analysis is not reported.

\section*{Funding information}

This research was funded by the EU H2020 FET-OPEN project NARCISO (No. 828890), the French
National Research Agency (ANR) through the projects
ULYSSES (No. ANR-15-CE24-0027-01) and OCTOPUS (No. ANR-18-CE47-0013-01), and funding from
the European Union - Next Generation EU (PNRR)
through the research project National Quantum Science
and Technology Institute (NQSTI).
M.B., F.I., and N.G. acknowledge the project PNRR MUR project PE0000023 - NQSTI. 
M.A. acknowledges the ANR AMPHORE project (ANR-21-CE09-0007) of the French Agence Nationale de la Recherche.
F.B. acknowledges Fondazione Cassa di Risparmio di Firenze for funding
this work within the projects Photonic Future
2021.1508 and PUPO (co-funded by the University of
Florence and the Italian Ministry of University and Research).

\section*{Acknowledgements}
We acknowledge the Nanotecmat platform of the IM2NP Institute of Marseille.

\section*{Conflict of interest}
The authors declare no conflict of interest

%\section*{Supporting Information}

%\printendnotes

%\bibliographystyle{...}
\bibliography{main}
% other bib styles:
% 1) Do not use either num-refs or alpha-refs in documentclass.
% 2) Load natbib package with the options set as needed.
% 3) Use the \bibliographystyle command to specify the style
% Included NJD styles are: 
%   WileyNJD-ACS
%   WileyNJD-AMA
%   WileyNJD-AMS
%   WileyNJD-APA
%   WileyNJD-Harvard
%   WileyNJD-VANCOUVER

%\graphicalabstract{example-image-1x1}{Please check the %journal's author guildines for whether a graphical %abstract, key points, new findings, or other items are %required for display in the Table of Contents.}

\newpage

\section*{Supplemental Information}
\setcounter{equation}{0}
\setcounter{figure}{0}
\setcounter{table}{0}
\setcounter{page}{1}
\setcounter{section}{0}

\renewcommand{\theequation}{S\arabic{equation}}
\renewcommand{\thefigure}{S\arabic{figure}}
\renewcommand{\thesection}{S\arabic{section}}
\renewcommand{\bibnumfmt}[1]{[S#1]}
\renewcommand{\vec}[1]{\boldsymbol{#1}}

\subsection{\label{sec:Raman} Raman spectroscopy of SiN/SOI patches}

\begin{figure}[h!]
\centering
\includegraphics[width=\columnwidth]{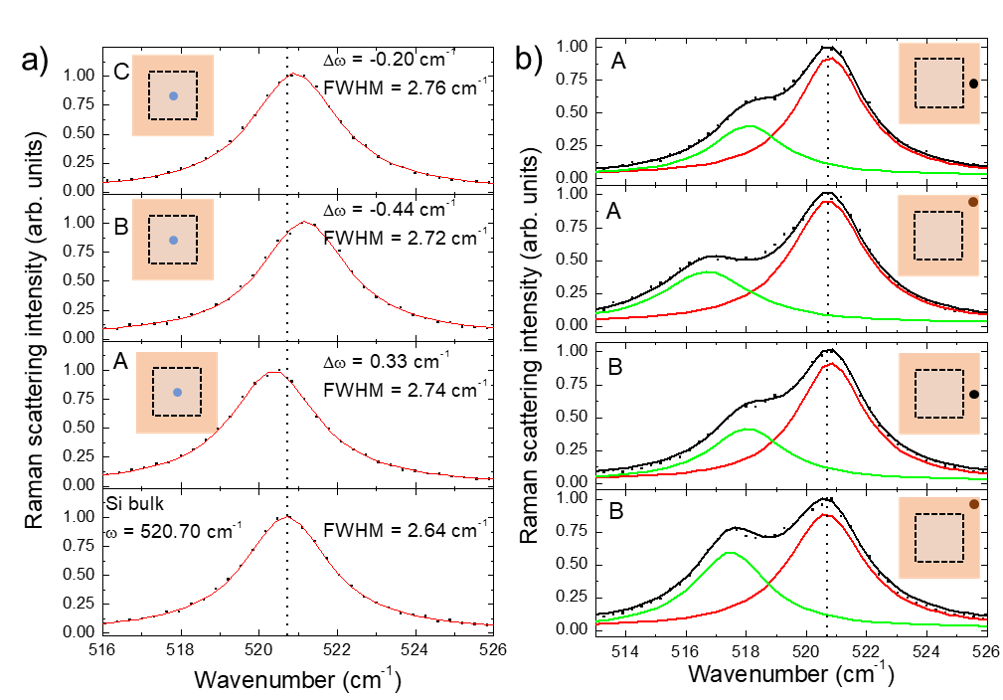}
\caption{ \textit{Raman spectroscopy of SOI patches}. a) From the bottom to the top panel are displayed Raman spectra of Si bulk, the center of the patch from samples A, B, and C. The experimental data are shown as square dots. The red curves are Lorentzian fits the data. The vertical, dotted line highlights the spectral position of the Raman signal of the bulk Si. The insets show the scheme of the membrane with a black dot highlighting the position of the detection. b) Raman spectra from samples A and B (respectively top panels and bottom panels) detected in the suspended parts, at the corner of the patch, and at the side. The experimental data are shown as square dots. Red and green curves represent Lorentzian fits to the data. The vertical, dotted lines highlight the spectral position of the Raman signal of bulk Si. The insets show the scheme of the membrane with a dot highlighting the position of the detection.}
\label{fig:SIfigureRAMAN}
\end{figure}

Raman spectroscopy is performed at room temperature on samples grown in the same conditions as those investigated in micro-photoluminescence, on patches having a side of 50~$\mu$m. They are summarized as follows (Figure~\ref{fig:SIfigureRAMAN}): i) Irrespective of the conditions used for SiN deposition (Table~\ref{tab:table1}), in the flat areas (far from the etched patches), the spectral shift of the SOI Raman signal is within the resolution of the experimental setup. As such, in these parts, the strain in the SiN is not relaxed in the SOI that is considered unstrained (not shown~\cite{ogura2009}). ii) At the center of the patches the Raman signal shows a peak broadening very close to that detected on Si (about 2.7~cm$^{-1}$). In this position, for SiN grown with plasma HF$\%$ = 100 and 70 (samples B and C), the strain is tensile, whereas, for plasma HF$\%$ = 0 (sample A), the strain is compressive. This information is consistent with high-resolution SEM that shows the bending of the membranes upwards or downwards, respectively for tensile or compressive strain (e.g. see Figure~\ref{fig:figure1} (c)). For these patches with 50~$\mu$m side, assuming an uniaxial stress, $\sigma$, can be written as $\sigma$(MPa) = -434 $\Delta \omega$(cm$^{-1}$) 
%[195, 196],
where $ \Delta \omega$ is the shift of the Raman peak with respect to the Si peak~\cite{bollani2015,barget2016}. For samples A, B, and C we find, respectively, a stress of about -143~MPa, 191~MPa, and 87~MPa (Figure~\ref{fig:SIfigureRAMAN} (a)). iii)  Raman spectra detected at the edge and at the corner of the membrane, in its suspended part, show a splitting of the peak in two components, having similar intensity and broadening (FWHM $\sim$ 2.8~cm$^{-1}$) and split of 2.5 to 4.5~cm$^{-1}$ (Figure~\ref{fig:figure2} (b)). This can be tentatively interpreted as the presence of a biaxial strain and the onset of a TO phonons associated with strain in the [110] direction\cite{urena2013,suess2014}.         

%\newpage

\subsection{\label{sec:Flat} Spectroscopy of flat SiN/SOI}

\begin{figure}[h!]
\centering
\includegraphics[width=\columnwidth]{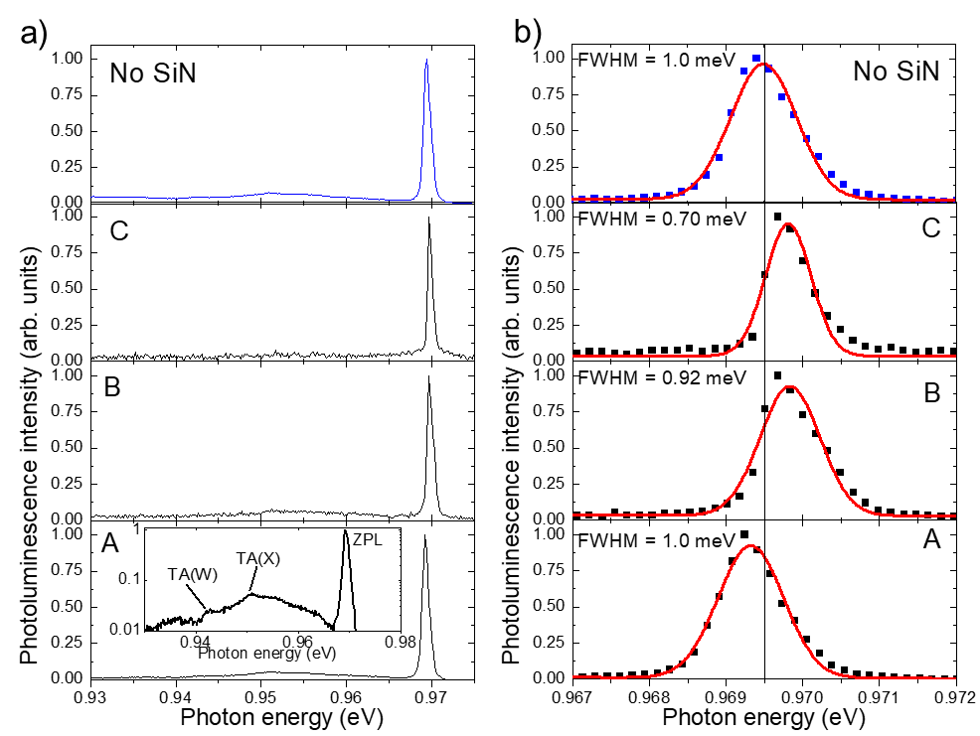}
\caption{ \textit{G-centers emission in flat areas}. a) Photoluminescence spectra collected in flat areas for samples A, B, C, and without SiN (from the bottom to the top panel, respectively). The inset in the bottom panel shows the G-center spectrum of sample A in the flat area in a logarithmic scale in order to highlight the presence of the TA(X) and TA(W) phonon lines. b) Blow up of the ZPL for samples A, B, C, and without SiN  (from bottom to top panel, respectively). The experimental data are represented as square dots whereas the red line is a Gaussian fit. The vertical line highlights the position of the ZPL for the un-strained case.}
\label{fig:figureFLAT}
\end{figure}

We characterize the emission of G-centers from flat areas for strained and unstrained samples (Figure~\ref{fig:figureFLAT}). We observe the typical spectrum composed of a sharp ZPL (at about 0.969 eV) and a broad phonon sideband. This latter contribution is composite and we can identify the presence of the TA(X) and TA(W) phonon modes~\cite{beaufils2018} (Figure~\ref{fig:figureFLAT} (a), inset of the bottom panel). The ZPLs display a Gaussian lineshape to a good approximation, small shifts (of about 0.2~meV, within the spectral resolution) with respect to the unstrained case and a similar broadening (0.7 meV to 1.0~meV). These observations highlight the negligible impact of the SiN deposition on the G-center dynamics in the flat areas.

\end{document}